\documentclass[]{mn2e}
\usepackage{graphicx,lscape}

\newif\ifAMStwofonts

\def\gs{\mathrel{\raise0.35ex\hbox{$\scriptstyle >$}\kern-0.6em 
\lower0.40ex\hbox{$\scriptstyle \sim$}}}
\def\ls{\mathrel{\raise0.35ex\hbox{$\scriptstyle <$}\kern-0.6em 
\lower0.40ex\hbox{$\scriptstyle \sim$}}}
\newcommand{\msun}{{\rm M}_\odot}

\newcommand{\be}{\begin{equation}}
\newcommand{\ee}{\end{equation}}
\newcommand{\bea}{\begin{eqnarray}}
\newcommand{\eea}{\end{eqnarray}}

\ifCUPmtlplainloaded \else
  \ifAMStwofonts \else 

  \fi
\fi

\title{Substructure in lensing clusters and simulations}
\author[Natarajan, De Lucia \& Springel]{Priyamvada Natarajan$^{1,2}$, 
Gabriella De Lucia$^3$ and Volker Springel$^3$\\
$^1$ Department of Astronomy, Yale University, P. O. Box 208101, 
New Haven, CT, 06511-208101, USA\\
$^2$ Department of Physics, Yale University, P. O. Box 208120, 
New Haven, CT 06511-208120, USA\\
$^3$ Max-Planck-Institut fur Astrophysik, Karl-Schwarzschild-Strasse 1, 
Postfach 1317, D-85748 Garching bei Munchen, Germany}

\pagerange{\pageref{firstpage}--\pageref{lastpage}}
\pubyear{2006}
 
\begin{document}
 
\maketitle
 
\label{firstpage}
 
\begin{abstract}
  
  We present high-resolution mass reconstructions for five massive
  cluster-lenses spanning a redshift range from $z = 0.18$--$0.57$
  utilising archival {\it Hubble Space Telescope} ({\it HST}) data and
  applying galaxy-galaxy lensing techniques.  These detailed mass
  models were obtained from the observations by combining constraints
  from the strong and weak lensing regimes. We ascribe local
  weak distortions in the shear maps to perturbations induced by the
  presence of galaxy haloes around individual bright early-type
  cluster member galaxies. This technique constrains the mass enclosed
  within an aperture for these subhaloes.  We are sensitive to a
  specific mass range for these subhaloes, $10^{11}$ --
  $10^{12.5}\,\msun$, which we associate with galaxy-scale
  subhaloes. Adopting a parametric model for the subhaloes, we also
  derive their velocity dispersion function and the aperture radius
  function. The mass spectrum of substructure in the inner regions of
  the observed clusters is directly compared with that in simulated
  clusters extracted from the {\it Millennium Simulation}.  The mass
  function, aperture radii and velocity dispersion function are
  compared in detail.  Overall, we find good agreement between the
  distribution of substructure properties retrieved using the lensing
  analysis and those obtained from the simulation.  We find that the
  fraction of total cluster mass associated with individual subhaloes
  within the inner 0.5 -- $0.8\,h^{-1}\,{\rm Mpc}$ of our clusters
  ranges from $10$--$20$ per cent, in broad agreement with
  simulations. Our work provides a powerful test of the $\Lambda$CDM
  model, which appears consistent with the amount of observed
  substructure in massive, lensing clusters based on present data.
\end{abstract}

\begin{keywords}
gravitational lensing, galaxies: fundamental parameters, haloes, methods:
numerical 
\end{keywords}

\section{Introduction}

Gravitational lensing has emerged as one of the most powerful
techniques to map mass distributions on a range of scales: galaxies,
clusters and beyond. The distortion in the shapes of background
galaxies viewed through foreground mass distributions is independent
of the dynamical state of the lens, therefore, compared with other methods
for mass estimation there are fewer biases in lensing mass
determinations. Here, we focus on mapping in detail the mass
distribution inside the inner regions of massive clusters of galaxies
using Hubble Space Telescope (HST) observations. We exploit the
technique of galaxy-galaxy lensing, which was originally proposed as a
method to constrain the masses and spatial extents of field galaxies
(Brainerd, Blandford \& Smail 1996), which we have since extended and
developed over the years to apply inside clusters (Natarajan \& Kneib
1997; Natarajan et al. 1998; 2002a).

The detailed mass distribution within clusters and specifically the
fraction of the total cluster mass associated with individual galaxies
has important implications for the frequency and nature of galaxy
interactions in clusters (Merritt 1983; Richstone 1976; Farouki \&
Shapiro 1981; Moore et al.\ 1996; Ghigna et al.\ 1998; Okamato \& Habe
1999).  Knowledge of the dynamical history of clusters enables a
deeper understanding of the physical processes that shape their
assembly and evolution.  The discovery of strong evolution between
$z\sim 0.5$ and the present-day in the morphological (and
star-formation) properties of the galaxy populations in clusters has
focused interest on environmental processes which could effect the
gaseous component and dark matter halo of a cluster galaxy (e.g.\
Couch et al.\ 1994, 1998).

The global tidal field of a massive, dense cluster potential well is expected
to be strong enough to truncate the dark matter halo of a galaxy whose orbit
penetrates the cluster core. Therefore, probing the extents of galaxy haloes in
clusters can provide invaluable clues to dynamically dominant processes in
clusters. For instance, the survival of individual, compact dark haloes
associated with cluster galaxies suggests a high probability for galaxy--galaxy
collisions within rich clusters over a Hubble time.  However, since the
internal velocity dispersions of cluster galaxies ($\ls 200$\,{\rm
  km\,s}$^{-1}$) are significantly lower than their orbital velocities, these
interactions are, in general, unlikely to lead to mergers, but rather 
encounters of the kind simulated in the galaxy harassment picture by Moore et
al.\ (1996, 1998).

Previous work on galaxy-galaxy lensing in the moderate redshift field
has identified a signal associated with massive haloes around typical
field galaxies, extending to beyond\footnote{We adopt $h=H_0/100\,{\rm
km\,s}^{-1}\,{\rm Mpc}^{-1}=0.7$ and $\Omega_{\Lambda}=0.7$, and scale
other published results to this choice of parameters.} 100\,kpc (e.g.\
Brainerd, Blandford \& Smail 1996; Ebbels et al.\ 2000; Hudson et al.\
1998; Hoekstra et al.\ 2004). In particular, Hoekstra et al.\ (2004)
report the detection of a finite truncation radius of $185 \pm 30$ kpc
via weak lensing by galaxies based on imaging data from the
Red-Sequence Cluster Survey. Galaxy-galaxy lensing results from the
analysis of the Sloan Digital Sky Survey data (McKay et al. 2002;
Sheldon et al. 2004; Guzik \& Seljak 2002) have contributed to a
deeper understanding of the relation between mass and light. Similar
analysis of galaxies in the cores of rich clusters suggests that the
average mass-to-light ratio and spatial extent of the dark matter
haloes associated with morphologically classified early-type galaxies
in these regions may differ from those of field galaxies with
comparable luminosity (Natarajan et al.\ 1998, 2002a).  We find that
at a given luminosity, galaxies in clusters have more compact halo
sizes and lower masses (by a factor of 2--5) compared to their field
counter-parts. The mass-to-light ratios inferred for cluster galaxies
in the V-band are also lower than those of field galaxies with
comparable luminosity.  This is a strong indication of the effect of
the dense environment on the properties of dark matter haloes.

In this paper, we present a determination of the mass function and
other detailed properties of substructure in clusters using
galaxy-galaxy lensing techniques. A high resolution mass model tightly
constrained by strong and weak lensing observations is constructed
including individual cluster galaxies and their associated dark matter
haloes. These lensing models are new and incorporate recent observed
spectroscopic redshifts of several additional multiple images.  We
show that over a limited mass range we can successfully construct the
mass function of subhaloes inside clusters.  In earlier work, we
compared the mass function obtained from lensing with that extracted
from a massive cluster simulated at extremely high resolution
(Natarajan \& Springel 2004). In this work, we are able to compare
with a large ensemble of simulated clusters, significantly
strengthening the statistical significance of our results.

N-body simulations represent an indispensable tool for investigating
the non-linear growth of structures in its full geometrical
complexity.  The high numerical resolution achieved in recent years
has demonstrated that the cores of dark matter haloes that fall into a
larger system can survive for a relatively long time as
self-gravitating objects orbiting in the smooth dark matter background
of the system.  A wealth of dark matter substructures is now routinely
detected and studied with the aid of N-body simulations confirming
that the existence of substructures is a generic prediction of
hierarchical structure formation in Cold Dark Matter (CDM) models.

The subhalo mass function (i.e.~the abundance of dark matter
substructure as a function of mass) represents an important prediction
of hierarchical CDM structure formation models and has been subject of
intense studies since the `dwarf galaxy crisis' was identified (Moore
et al.\ 1999; Klypin et al.\ 1999; Stoehr et al. 2003).  Within a radius of
$400\,h^{-1}\,{\rm kpc}$, from the Milky Way, cosmological models of
structure formation predict $\sim$ 30 dark matter satellites with
circular velocities in excess of $20\,{\rm km\,s^{-1}}$ and mass
greater than $3\,\times\,10^8\,M_{\odot}$.  This number is
significantly higher than the dozen or so satellites actually observed
around our Galaxy.  Different explanations have been suggested for
this discrepancy.  The missing satellites could for example be
identified with the detected High Velocity Clouds (Maller \& Bullock
2004).  {\it Warm} or {\it self-interacting} dark matter could also
selectively suppress power on the small scales, therefore reducing the
predicted number of satellites (Spergel \& Steinhardt 2000). The
leading hypothesis however remains that the solution to this problem
lies in astrophysical processes such as heating by a photo-ionising
background that suppresses star formation in small haloes at early
times (Bullock et al.\ 2000; Benson et al.\ 2002; Kravtsov et a.\
2004).  On the scale of galaxy clusters, many more dark matter
structures are expected to host visible galaxies, thus making the
comparison with expectations from numerical simulations less affected
by uncertainties in the physics of galaxy formation.  Full consistency
is to be expected here, therefore providing an important test of the
CDM paradigm.

Our innovative application of gravitational lensing enables a mapping
of the distribution in mass of dark matter substructure (subhalo mass
function) therefore allowing a direct comparison with results from
numerical simulations. The strength of the lensing analysis presented
here derives from the combination of both strong and weak lensing
features which are used together to construct a high resolution mass
map of a galaxy cluster. Anisotropies in the shear field (i.e.\
departure from the coherent tangential signal) in the vicinity of
bright, early-type cluster members are attributed to the presence of
these local potential wells. Statistically stacking this signal
provides a way to quantify the masses associated with individual
galaxy haloes. This is accomplished using a maximum likelihood
estimator to retrieve characteristic properties for a typical subhalo
in the cluster.

The comparison with clusters from the {\it Millennium Simulation} we
use here is particular powerful due to the large volume and high
resolution of this simulation. The simulation used more than 10
billion particles to trace the evolution of the dark matter
distribution in a cubic region of the Universe over 2 billion
light-years on a side.  This makes it an ideal data set for comparison
with lensing clusters, providing dozens of highly resolved, massive
lensing clusters out to redshifts $z \simeq 0.6$ and beyond.

The outline of this paper is as follows: in Section~2, we describe
briefly the formalism for analysing galaxy-galaxy lensing in observed
clusters including a synopsis of the adopted models. Section~3
presents the best-fit lens models and discusses the uncertainties and
sources of error.  A detailed comparison with clusters from the
Millennium Simulation is presented in Section~4. The results of
comparing the mass function of substructure, the distributions of
aperture radii and the velocity dispersions are discussed. Finally,
we conclude with a discussion of the implications of our results for
the $\Lambda$CDM model and the future prospects of this work in
Section~5.
\section{Galaxy-galaxy lensing in clusters}

\subsection{Framework for analysis}

In this section, we briefly outline the analysis framework, noting
that further details can be found in earlier papers (Natarajan \&
Kneib 1997; Natarajan et al. 1998).  For the purpose of extracting the
properties of the subhalo population in clusters, a range of mass
scales is modelled parametrically. The X-ray surface brightness maps
of clusters suggest the presence of a smooth, dominant, large scale
mass component. Clusters are therefore modelled as a super-position of
a smooth large-scale potential and smaller scale potentials that are
associated with bright early-type cluster members:
\begin{equation}
\phi_{\rm tot} = \phi_{\rm smooth} + \Sigma_i \,\phi_{\rm p_i},
\end{equation}
where $\phi_{\rm smooth}$ is the potential of the smooth component and 
$\phi_{\rm  p_i}$ is the potential of the subhalo associated with the $i$th
galaxy, and is treated as perturber.  The  amplification matrix $A^{-1}$ can 
be decomposed into contributions from the main clump and the perturbers:
\bea 
A^{-1}\,=\,(1\,-\,\kappa_{\rm smooth}\,-\,\Sigma_i \kappa_{\rm
p})\,I - \gamma_{\rm smooth}J_{2\theta_{\rm smooth}} - \Sigma_i \,\gamma_{\rm
p_i}J_{2\theta_{\rm p_i}} \nonumber, 
\eea 
where $\kappa$ is the magnification
and $\gamma$ the shear.  The shear $\gamma$ is in fact a complex
number and is used to define the reduced shear $\overline{g}$, 
which is the quantity that is measured directly from the observed 
shapes of background galaxies. Similarly, the reduced shear
can also be decomposed as
\bea
\overline{g_{\rm tot}} = {\overline{\gamma} \over 1-\kappa} =
{{\overline\gamma_{\rm smooth}} + \Sigma_i \,{\overline\gamma_{p_i}} \over
1-\kappa_{\rm smooth} -\Sigma_i \,\kappa_{p_i}}.
\eea 
In the frame of an individual perturber $j$ (and neglecting the effect 
of perturber $i$ if $i \neq j$), this simplifies to: 
\bea 
{\overline g_{\rm tot}}|_j} = { {{\overline \gamma_{\rm smooth}}
+{\overline \gamma_{p_j}} \over {1-\kappa_{\rm smooth} -\kappa_{p_j }}}. 
\eea
Restricting our analysis to the weak regime, and thereby retaining
only the first order terms from the lensing equation for the shape
parameters (e.g. Kneib et al.\ 1996), we have: 

\be 
{\overline g_I}=
{\overline g_S}+{\overline g_{\rm tot}}, 
\ee 
where ${\overline g_I}$\footnote{The
measured image shape and orientation are used to construct a complex number
whose magnitude is given in terms of the semi-major axis (a) and
semi-minor axis (b) of the image and the orientation is the phase of
the complex number.} is the distortion 
of the image, ${\overline g_S}$ the intrinsic shape of the source, and
${\overline g_{\rm tot}}$
is the distortion induced by the lensing potentials.

In the local frame of reference of the perturbers, the mean value of
the quantity ${\overline g_I}$ and its dispersion can be computed in
circular annuli (at radius $r$ from the perturber centre)
{\underline{strictly in the weak-regime}}, assuming a constant value
$\gamma_c e^{i\theta_{c}}$ for the smooth cluster component over the
area of integration. In the frame of the perturber, the averaging
procedure allows efficient subtraction of the large-scale component,
enabling the extraction of the shear component induced in the
background galaxies only by the local perturber. The background
galaxies are assumed to have intrinsic ellipticities drawn from a
known distribution (see the next section for further
details). Schematically, the effect of the cluster on the intrinsic
ellipticity distribution of background sources is to cause a coherent
displacement ${\tau}$ and the presence of perturbers merely adds
small-scale noise to the observed ellipticity distribution.

The feasibility and robustness of signal detection has been amply
demonstrated in earlier papers by Natarajan et al.  The primary
limitations in this analysis arise from the total number of distorted
background galaxies, and the accuracy with which the smooth cluster
component can be constrained using multiple images in the inner
region. The partition into this smooth component and its effective
subtraction in fact boosts the shear induced by the perturber. In
particular, the shear induced by the subhalo has a ($\kappa_{\rm
smooth}+\kappa_{p_j}$) term in the denominator, which becomes
non-negligible in the cluster centre.  The subtraction of the
larger-scale component reduces the noise in the polarisation measure
by about a factor of two in cluster cores. This differenced averaging
prescription for extracting the distortions induced by the possible
presence of dark haloes around cluster galaxies is feasible with HST
quality data as we have shown in earlier work (Natarajan et al.\ 1998,
2002a). The robustness of this technique also has been demonstrated in
our earlier published work. Note here that it is the presence of the
underlying large-scale smooth mass distribution (with a high value of
$\kappa_c$) that enables the extraction of the weak signal riding on
it. It is instructive to keep in mind that, in the regimes of interest
discussed here, the distortion induced by the cluster-scale smooth
component for a pseudo-isothermal elliptical component model (PIEMD)
in the inner-most (with a velocity dispersion of $1000\,{\rm
kms}^{-1}$ and at $R/r_t \leq 0.1$) regions is typically of the order
of 20 - 40 per cent or so in background galaxy shapes, and the
perturbers produce distortions (smaller scale PIEMDs with a velocity
dispersion of $220\,{\rm kms}^{-1}$) of the order of 5 - 10 per cent,
significantly more than in the case of weak-lensing by large scale
structure or cosmic shear, wherein the distortions are of the order of
1 per cent.

\subsection{Modelling the cluster}

Each of the clusters studied in this paper preferentially probes the
high mass end of the cluster mass function and has a surface
mass density in the inner regions which is higher than the critical
value, therefore producing a number of multiple images of background
sources. By definition, the critical surface mass density for strong lensing 
is given by:
\begin{eqnarray}
\Sigma_{\rm crit} = {\frac{c^2}{4 \pi G}} \frac{D_s}{D_d D_{ds}},
\end{eqnarray}
where $D_s$ is the angular diameter distance between the observer and the
source, $D_d$ the angular diameter distance between the observer and
the deflecting lens, and $D_{ds}$ the angular diameter distance between
the deflector and the source. When the surface mass density in the
cluster is in excess of this critical value, strong lensing phenomena
with high magnification are observed. 

In general two types of lensing effects are produced -- strong:
multiple images and highly distorted arcs; and weak: small distortions
in background image shapes determined by the criticality of the
region. Viewed through the central, dense core region of the mass
distribution, where $\kappa > 1$ strongly lensed features are
observed. Note that the integrated lensing signal detected is due to
all the mass distributed along the line of sight in a cylinder
projected onto the lens plane. In this and all other cluster lensing
work, the assumption is made that individual clusters dominate the
lensing signal as the probability of encountering two massive rich
clusters along the same line-of-sight is extremely small due to the
fact that these are very rare objects in hierarchical structure
formation models. 

With our current sensitivity limits, galaxy-galaxy lensing
within the cluster is primarily a tool to determine the total enclosed
mass within an aperture. We lack sufficient sensitivity to constrain
the detailed mass profile for individual cluster galaxies. With higher
resolution data in the near future we will be able to obtain
constraints on the slopes of mass profiles in subhaloes. In this
paper, we therefore concentrate on pseudo-isothermal elliptical
components (PIEMD models, derived by Kassiola \& Kovner 1993)
appropriately scaled for both the main cluster and the
substructures. Some of the cluster studied in this work are bi-modal (i.e.
have two significant, large-scale mass peaks), therefore, when required 
we employ two large-scale smooth potentials and a super-position of
subhaloes. We find that the results obtained for the
characteristics of the subhaloes (or perturbers) is largely
independent of the form of the mass distribution (the aperture mass 
is comparable for the NFW and PIEMD models) used to model the
smooth, large-scale component.  

To quantify the lensing distortion induced by the global potential,
both the smooth and individual galaxy-scale haloes are modelled
self-similarly using a surface density profile, $\Sigma(R)$, which is a
linear superposition of two PIEMD distributions,
\begin{eqnarray}
\Sigma(R)\,=\,{\Sigma_0 r_0  \over {1 - r_0/r_t}}
\left({1 \over \sqrt{r_0^2+R^2}}\,-\,{1 \over \sqrt{r_t^2+R^2}}\right),
\end{eqnarray}
with a model core-radius $r_0$ and a truncation radius $r_t\,\gg\,
r_0$. These
parameters $(r_t,r_0)$ are tuned for both the smooth component and the
perturbers to obtain mass distributions on the relevant scales. The
coordinate $R$ is a function of $x$, $y$ and the ellipticity, 
\bea
R^2\,=\,{x^2 \over (1+\epsilon)^2}\,+\,{y^2 \over
  (1-\epsilon)^2}, \;\;{\rm where}\;
\ \ \epsilon= {a-b \over a+b}.
\eea 
The mass enclosed within radius $R$ for the $\epsilon = 0$ model is given by 
\be 
M(R)={2\pi\Sigma_0
r_0 \over {1-{{r_0} \over {r_t}}}}
\left[\,\sqrt{r_0^2+R^2}\,-\,\sqrt{r_t^2+R^2}\,+\,(r_t-r_0)\,\right].  
\ee 
One of the attractive features of this
model is that the total mass $M$ is finite, $M\,
\propto \,{\Sigma_0} {r_0} {r_t}$. Besides, 
analytic expressions can be obtained for the all the quantities of
interest, $\kappa$, $\gamma$ and $g$, e.g.\bea
\kappa(R)\,=\,{\kappa_0}\,{{r_0} \over {(1 - {r_0/r_t})}}\, \left({1 \over
{\sqrt{{r_0^2}+{R^2}}}}\,-\,{1 \over
{\sqrt{{r_t^2}+{R^2}}}} \right)\,\,\,, \eea \bea
2\kappa_0\,=\,\Sigma_0\,{4\pi G \over c^2}\,{D_{\rm ls}D_{\rm ol}
\over D_{\rm os}}, \eea where $D_{\rm ls}$, $D_{\rm os}$ and $D_{\rm
ol}$ are the lens-source, observer-source and
observer-lens angular diameter distances,  respectively, 
which do depend on the choice
of cosmological parameters. To obtain $g(R)$ knowing the
magnification $\kappa(R)$, we solve Laplace's equation for the
projected potential $\phi_{\rm 2D}$, evaluate the components of the
amplification matrix and then proceed to solve directly for
$\gamma(R)$, and then $g(R)$:
\bea
\gamma(R)\,&=&\,\nonumber \kappa_0 \left[\,-{1 \over \sqrt{R^2 + r_0^2}}\, +\,{2 \over
R^2}(\sqrt{R^2 + r_0^2}-r_0)\,\right.\\ \nonumber &+&\left. \,{1 \over {\sqrt{R^2 +
r_t^2}}}\,-\, {2 \over R^2}(\sqrt{R^2 + r_t^2} - r_t)\,\right].\\ 
\eea
Scaling this relation by $r_t$ gives for $r_0<R<r_t$: 
\be
\gamma(R/r_t)\propto {\Sigma_0 \over \eta-1} {{r_t} \over
R}\,\sim\,{\sigma^2 \over R}, 
\ee 
where $\sigma$ is the velocity dispersion and for $r_0<r_t<R$: 
\be
\gamma(R/r_t)\propto {\Sigma_0\over\eta} {{r_t}^2 \over
R^2}\,\sim\,{{M_{\rm tot}} \over {R^2}}, 
\ee 
where ${M_{\rm tot}}$ is the total mass. In the limit that
$R\,\gg\,r_t$, we have 
\bea 
\gamma(R)\,=\,{{3 \kappa_{0}} \over {2
{R^3}}}\,[{r_{0}^2}\,-\,{r_{t}^2}]\,+\,{{2 {\kappa_0}} \over {R^2}}
[{{r_t}\,-\,{r_0}}], 
\eea 
and as ${R\,\to\,\infty}$, $\gamma(R)\,\to\,0$, $g(R)\,\to\,0$ and
$\tau(R)\,\to\,0$, as expected.

It is further assumed that the ellipticity and the orientation of the
dark matter subhaloes associated with the early-type cluster members
is identical to that of the galaxies themselves.  Additionally, in
order to relate the light distribution to key parameters of the mass
model above, we adopt a set of physically motivated scaling laws for
the cluster galaxies (Brainerd et al.\ 1996):
\begin{eqnarray}
{\sigma_0}\,=\,{\sigma_{0*}}\left({L \over L^*}\right)^{1 \over 4};\,\,
{r_0}\,=\,{r_{0*}}{\left({L \over L^*}\right)^{1 \over 2}};\,\,
{r_t}\,=\,{r_{t*}}{\left({L \over L^*}\right)^{\alpha}}.
\end{eqnarray}
These in turn imply the following scaling for the $r_t/r_0$ ratio $\eta$:
\bea
{\eta}\,=\,{r_t\over r_0}={{r_{t*}} \over {r_{0*}}} 
\left({L \over L^*}\right)^{\alpha-1/2}.
\eea
The total mass $M_{\rm ap}$ enclosed within an aperture $r_{t*}$ and
the total mass-to-light ratio $M/L$ 
then scale with the luminosity as follows:
\begin{eqnarray}
M_{\rm ap}\,\propto\,{\sigma_{0*}^2}{r_{t*}}\,\left({L \over L^*}\right)^{{1 \over
2}+\alpha},\,\,{M/L}\,\propto\,
{\sigma_{0*}^2}\,{r_{t*}}\left( {L \over L^*} \right)^{\alpha-1/2},
\end{eqnarray}
where $\alpha$ determines the size of the galaxy halo. For $\alpha$ = 0.5
the assumed galaxy model has constant $M/L$ with luminosity (but not
as a function of radius) for each galaxy. We adopt $\alpha = 0.5$ throughout this 
work, as this scaling is empirically motivated by the
Faber-Jackson relation for early-type galaxies (Brainerd, Blandford \&
Smail 1996). We assume these scaling relations and recognise that this
could ultimately be a limitation of our model but the evidence at hand
supports the fact that mass traces light efficiently both on cluster
scales (Kneib et al. 2003) and on galaxy scales (McKay et al. 2002;
Wilson et al. 2001). The calibration factors $\sigma_{0*}$ and $r_{t*}$ 
in eqn. (15) are provided by the results of a likelihood method discussed
further in Section 2.3.  

\subsubsection{The intrinsic shape distribution of background galaxies}

As in all lensing work, it is assumed here as well that the intrinsic or
undistorted distribution of shapes of background galaxies is known. This
distribution is obtained from shape measurements taken from deep
images of blank field surveys. Previous analysis of deep survey
data such as the MDS fields (Griffiths et al.\ 1994) showed that the 
ellipticity distribution of sources is a strong function of the sizes 
of individual galaxies as well as of their magnitude (Kneib et al.\
1996). For the purposes of our modelling, the intrinsic ellipticities
for background galaxies are assigned in concordance with an
ellipticity distribution $p(\tau_S)$ where the shape parameter 
$\tau$ is defined as $\tau = (a^2-b^2)/(2ab)$ derived from the observed
ellipticities of the CFHT12k data (see Limousin et al.\ 2005 for details): 
\be
p(\tau_S)\,=\,\tau_S\,\,\exp\left[-\left({\tau_S \over
\delta}\right)^{\nu}\right];\,\,\,\nu\,=\,1.15,\,\,\delta\,=\,0.25.  
\ee
Note that this distribution includes accurately measured shapes of
galaxies of all morphological types. In the likelihood analysis
this distribution $p(\tau_S)$ is the assumed prior, which is used to
compare with the observed shapes once the effects of the assumed 
mass model are removed from the background images. We note here that 
the exact shape of the ellipticity distribution, i.e. the functional 
form and the value of $\delta$ and $\nu$ do not change the results, 
but alter the confidence levels we obtain. The width of the intrinsic
ellipticity distribution on the other hand is the fundamental
limiting factor in the accuracy of all lensing measurements.

\subsubsection{The redshift distribution of background galaxies}

While the shapes of lensed background galaxies can be measured
directly and reliably by extracting the second moment of the light
distribution, the precise redshift for each weakly object
is in general unknown and therefore needs to be assumed. Using
multi-waveband data from surveys such as COMBO-17 (Wolf et al.\ 2004),
photometric redshift estimates can be obtained for every background
object. Typically the redshift distribution of background galaxies is modelled
as a function of observed magnitude $P(z,m)$. We have used data from
the high-redshift survey VIMOS VLT Deep Survey (Le Fevre et al.
2004) as well as recent CFHT12k R-band data to define the number counts of
galaxies, and the HDF prescription for the mean redshift per magnitude
bin, and find that the simple parameterisation of the redshift
distribution used by Brainerd, Blandford \& Smail (1996) still
provides a good description to the data.

For the normalised redshift distribution at a given magnitude $m$ (in
the given band) we therefore have \bea N(z)|_{m}\,=\,{{\beta\,({{z^2}
\over {z_0^2}})\, \exp[-({z \over {z_0}})^{\beta}]} \over {\Gamma({3
\over \beta})\,{{z_0}}}}, \eea where $\beta\,=\,$1.5 and \bea
z_0\,=\,0.7\,\left[\,{z_{\rm median}}\,+\,{{{\rm d}{z_{\rm median}}}
\over {{\rm d}m_R}}{(m_R\,-\,{m_{R0}})}\,\right], \eea with ${z_{\rm
median}}$ being the median redshift, and ${\rm d}z_{\rm median}/ {\rm
d}m_R$ being the change in median redshift with say the $R$-band
magnitude, $m_R$.

However, we note here in agreement with another recent study of
galaxy-galaxy lensing in the field by Kleinheinrich et al. (2005),
that the final results for the aperture mass presented here are
primarily sensitive to the choice of the median redshift of the
distribution rather than the individual assigned values.

\subsection{The maximum-likelihood method}

There are two key aspects to constructing a successful lens model for the
clusters analyzed here. Firstly, we must identify multiple-imaged
background sources with reliable redshift measurements whose
properties can be used to constrain the total projected mass within
the Einstein radius. The strong lensing observations provide constraints
in the inner regions for the global smooth potential. 
Secondly, we use the weak shear detected out to larger radii to 
constrain the behavior of the smooth component and simultaneously 
statistically constrain the properties of the subhaloes. This is done
by quantifying the anisotropies that arise in the shear field and
attributing these anisotropies to the presence finite mass subhaloes.

Parameters that characterise both the global component and the
perturbers are optimised, using the observed strong lensing features -
positions, magnitudes, geometry of multiple images and measured
spectroscopic redshifts, when known, along with the smoothed shear
field as constraints. As initial values of the mass model we provide a
center, velocity dispersion, ellipticity, orientation, and truncation
radius for the main clump with tolerance ranges. For the subhaloes, we
provide the locations (coincident with the positions of the selected
early-type cluster members), ellipticity and orientation (taken from
measurements for the cluster early-types). In combination with the
parameterisation presented in the previous section, we then optimise
and extract values for the central velocity dispersion and the
aperture scale $(\sigma_{0*}, r_{t*})$ for a typical $L^*$ cluster
galaxy.

A maximum-likelihood estimator is used to obtain significance bounds
on fiducial parameters that characterise a typical $L^*$ subhalo in
the cluster. We have extended the prescription proposed by Schneider
\& Rix (1997) for galaxy-galaxy lensing in the field to the case of
lensing by galaxy subhaloes in the cluster (Natarajan \& Kneib 1997,
Natarajan et al 1998). The likelihood function of the estimated
probability distribution of the source ellipticities is maximised for
a set of model parameters, given a functional form of the intrinsic
ellipticity distribution measured for faint galaxies.  For each
`faint' galaxy $j$, with measured shape $\tau_{\rm obs}$, the
intrinsic shape $\tau_{S_j}$ is estimated in the weak regime by
subtracting the lensing distortion induced by the smooth cluster model
and the galaxy subhaloes,
\begin{eqnarray}
\tau_{S_j} \,=\,\tau_{\rm obs_j}\,-{\Sigma_i^{N_c}}\,
{\gamma_{p_i}}\,-\, \gamma_{c}, 
\end{eqnarray}
where $\Sigma_{i}^{N_{c}}\,{\gamma_{p_i}}$ is the sum of the shear
contribution at a given position $j$ from $N_{c}$ perturbers. This
entire inversion procedure is performed numerically using code that
builds on the ray-tracing routine {\sc lenstool} written by Kneib
(1993).  This machinery accurately takes into account the
non-linearities arising in the strong lensing regime. Using a
well-determined `strong lensing' model for the inner regions of the
clusters derived from the positions, shapes and magnitudes of the
highly distorted multiple-imaged objects along with the shear field
determined from the shapes of the weakly distorted background galaxies
and assuming a known functional form for $p(\tau_{S})$, the probability
distribution for the intrinsic shape distribution of galaxies in the
field, the likelihood for a guessed model is given by
\begin{eqnarray}
 {\cal L}({{\sigma_{0*}}},{r_{t*}}) = 
\large\Pi_j^{N_{\rm gal}} p(\tau_{S_j}),
\end{eqnarray}
where the marginalisation is done over $(\sigma_{0*},r_{t*})$.  We
compute ${\cal L}$ assigning the median redshift corresponding to the
observed source magnitude for each arclet. The best fitting model
parameters are then obtained by maximising the log-likelihood function
with respect to the parameters ${\sigma_{0*}}$ and ${r_{t*}}$.  Note
that the parameters that characterise the smooth component are also
simultaneously optimised. The likelihood can also be marginalised over
a complementary pair of parameters, e.g.~using the luminosity scaling
index $\alpha$ and the aperture mass $M_{\rm ap}$ directly. In this
work, we explore both choices.

\section{Best-fit lensing mass models}

A composite mass model is constructed for the clusters starting with
the super-posed PIEMDs. The strong lensing data, i.e. the geometry,
positions, relative brightness, redshifts and parities of the multiple
images are used to obtain the mass enclosed within the Einstein radius which
is used as an initial constraint for the integrated mass in the inner
regions. The contribution to the shear and magnification from all
potentials (large-scale and small-scale perturbers) is calculated at
the location of every observed background source galaxy and the
inversion of the lensing equation is performed. The observed shape and
magnification of each and every distorted background galaxy is
compared to that computed from the model and the subhalo mass
distribution is modified iteratively until the best match between the
observations and the model is found simultaneously for all background
sources. Strong lensing constraints principally drive the likelihood
results. Therefore, we use many sets of multiple images with measured
redshifts for each cluster as inputs. Compared to our earlier work 
(Natarajan et al. 2002) the models presented here have several improvements,
we have incorporated more multiple image families with redshifts that
have since become available and we have also modified the algorithm
to enable finer grid searching. In our analysis, the center of the cluster is
picked to coincide with the location of the brightest cluster galaxy.

The basic steps of our analysis involve lens inversion, modeling and
optimization, which are done using the {\sc lenstool} software
utilities (Kneib 1993). These utilities are used to perform the ray
tracing from the image plane to the source plane with a specified
intervening lens. This is achieved by solving the lens equation
iteratively, taking into account the observed strong lensing features,
positions, geometries and magnitudes of the multiple images.  In some
cases, we also include a constraint on the location of the critical
line to tighten the optimization.  Additionally, we fix the core
radius of an $L^*$ subhalo to be $0.1\,{\rm kpc}$, as by construction
our analysis cannot constrain this quantity. In addition to the
likelihood contours, the reduced $\chi^2$ for the best-fit model is
also robust. We describe some pertinent features of each cluster and
their respective mass models below.

\begin{table*}
\begin{center}
\begin{tabular}{lcccccccc}
\hline\hline\noalign{\smallskip} ${\rm Cluster}$&{$z$} &
${\sigma^\ast}$&${r_t^\ast}$&${M_{\rm ap}/L_v}$&$ {M^\ast}$&
$\sigma_{\rm clus}$ & ${\rho_{\rm clus}(r = 0)}$\\ & & (km\,s$^{-1}$)
& (kpc) & (M$_\odot$/L$_\odot$) & (10$^{11}$M$_\odot$) &
(km\,s$^{-1}$) & (10$^6$ $\msun$ kpc$^{-3}$)\\ \noalign{\smallskip}
\hline \noalign{\smallskip} {A\,2218} & ${0.17}$ & ${180\pm10}$ &
${40\pm12}$ & ${5.8\pm1.5}$ & $\sim\,14 $ & ${1070\pm70}$ & {3.95}\\

{A\,2390} & ${0.23}$ & ${200\pm15}$ & ${18\pm5}$ &A
${4.2\pm1.3}$ & $\sim\,6.4 $  &${1100\pm80}$& {16.95}\\

{AC\,114}  & ${0.31}$ & ${192\pm35}$ & ${17\pm5}$ &
${6.2\pm1.4}$ & $\sim\,4.9 $ &${950\pm50}$& {9.12}\\ 

{Cl\,2244$-$02} & ${0.33}$ & ${110\pm7}$ & ${55\pm12}$ &
${3.2\pm1.2}$ & $\sim\,6.8 $ &${600\pm80}$  & {3.52}\\

{Cl\,0024+16} & ${0.39}$ & ${125\pm7}$ & ${45\pm5}$ &
${2.5\pm1.2}$ & $\sim\,6.3 $ &${1000\pm70}$ & {3.63}\\

{Cl\,0054$-$27} & ${0.57}$& ${230\pm18}$ & ${20\pm7}$ &
${5.2\pm1.4}$ & $\sim\,9.4 $ &${1100\pm100}$ & {15.84}\\
\noalign{\smallskip}
\hline
\end{tabular}
\end{center}
\caption{Parameters that define the mass models of the subhaloes
for the lensing clusters.}
\end{table*}

\noindent{\bf A\,2218}

Our best fit mass model for the cluster is bimodal, composed of two
large scale clumps around the cD and the second brightest cluster
galaxy.  This model is an updated version of that
constructed by Kneib et al.\ (1996). It includes 40 additional
small-scale clumps that we associate with luminous early-type galaxies
in the cluster core. Only about 10 per cent of the total cluster mass is in
substructures, i.e.\ associated with galaxy scale haloes. The aperture
mass, integrated over the truncation radius $r_{\rm ap}\,=\,40$ kpc,
yields a characteristic mass of $1.4 \times 10^{12}\,\msun$, with a
total mass-to-light ratio in the V-band of $\sim\,5.8 \pm 1.5$ and a central
velocity dispersion of about $180\,{\rm km\,s}^{-1}$. 

\noindent{\bf A\,2390}

The cluster has an unusual feature -- a strongly lensed almost
`straight arc' (Pello et al.\ 1991) approximately 38 arcsec ($\sim$
170 kpc) away from the central galaxy, in addition to many other arcs
and arclets that have been utilised in our modeling. We find a
best-fit mass model with two large-scale components, that yield a
projected mass within the radius defined by the brightest arc of $\sim
1.8 \pm 0.2 \times 10^{14}\,M_\odot$.  Our best-fit composite lensing
model for A\,2390 incorporates 40 perturbers associated with
early-type cluster members whose characteristic parameters are
optimized in the maximum-likelihood analysis. The integrated mass
within the $\sim$ 18 kpc tidal radius for a typical $L^*$ cluster
galaxy is about $6.4 \times 10^{11}\,\msun$ giving a total
mass-to-light ratio in the V-band of about $4.2 \pm 1.3$. Again, 90
per cent of the total mass of the cluster is consistent with being
smoothly distributed.

\noindent{\bf Cl\,2244$-$02}

The best-fit lens model for this cluster has two components which both
have fairly low velocity dispersions. This is the least massive
lensing cluster in the sample studied here. The X-ray mass estimate
from the {\it ASCA} data (Ota et al.\ 1998) is in good agreement with
our best-fit lensing mass model. This is despite the fact that the
X-ray temperature of Cl\,2244$-$02 is at least a factor of two higher
than that expected from the average luminosity-temperature relation.

The tidal truncation radius obtained for a typical $L^*$ cluster
galaxy in Cl\,2244 is the largest in the sample studied here and is
$55 \pm 12$ kpc. This is consistent with the fact that the central
density in Cl\,2244 is the lowest. The total mass-to-light ratio in
the V-band for a fiducial $L^*$ is $3.2 \pm 1.2$. Approximately 20 per cent
of the total mass is in substructure within the mass range $10^{11} -
10^{12.5}\,\msun$.

\noindent{\bf Cl\,0024+16}

Our best fit mass model for the inner regions takes into account the
small scale dark haloes associated with the early-type members in the
core, and requires a two component model for the sub-clusters.
Integrating the best-fit mass model shown, we find that (i) about 10
per cent of the total cluster mass is in galaxy-scale haloes and (ii)
the total mass estimate is in good agreement with that obtained by
Kneib et al.\ (2003) where data from a much larger field of view were
used.

Even on the large scales probed by Kneib et al.\ (2003) it was found
that mass and light traced each other rather well at large
radii. A typical $L^*$
cluster galaxy was found to have  a truncation radius of $45 \pm 5$ kpc, and a
central velocity dispersion of $125 \pm 7\,{\rm km\,s}^{-1}$.

\noindent{\bf Cl\,0054$-$27}

The lensing signal from Cl\,0054$-$27 is best fit by a single smooth dark
matter component and subhaloes associated with bright, early-type members
making it the only uni-modal cluster in the sample studied here. The mass
enclosed within $\sim$ 400 kpc is of the order of $1.8 \pm 0.4 \times
10^{14}\,\odot$.

The characteristic central velocity dispersion of a typical $L^*$
galaxy in this cluster is higher than in A\,2218, A\,2390 or
Cl\,0024+16, all of which are by contrast bimodal in the mass
distribution. In this cluster, about 20 per cent of the total mass is
in substructure. However, Cl\,0054$-$27 is the most distant cluster
studied here and is likely to be still evolving and assembling,
accounting for the high mass fraction in substructure.

Note here that the choice of $\alpha$ determines only the scaling of
the outer radius of a fiducial subhalo with luminosity.  With the data
used in this paper it is not possible to distinguish between various
values of $\alpha$ - some values are clearly more physical than
others.  Therefore, this implies that we are sensitive to the
integrated mass within an aperture that is determined primarily by the
anisotropy in the shear field and not by the details of how the
subhalo masses are truncated. We also find that out to 500 kpc in all
clusters only 10--20 per cent of the total mass is associated with
galaxy haloes. At this radius (to which we are limited due to the size
of the HST WFPC2 fields), most of the mass is in the large-scale
component. This fraction is likely to be a strong function of
cluster-centric radius. The dependence of these aperture radii with
distance from the cluster can be explored with wide-field HST data and
we are in the process of doing so for the cluster Cl\,0024+16
(Natarajan et al.\ 2006).

\begin{table*}[ht!] 
\begin{center}
\begin{tabular}{lccccccc}
\hline\hline\noalign{\smallskip}
$z$& $x$ & $y$ & $\epsilon$ & ${\theta}$ & ${\sigma}$ & ${r_t}$ &$ {r_c}$\\
${}$&${\rm arcsec}$&${\rm arcsec}$&${}$&${\rm deg}$&$(km\,s^{-1})$ &$(\rm kpc)$&
$(\rm kpc)$\\
\noalign{\smallskip}
\hline
 \noalign{\smallskip}
{\bf A\,2218}\\
${0.17}$ & ${0.5}$ & ${1.2}$ & ${0.2}$ & ${-17}$ & ${1100}$ & ${900}$ 
& ${65}$\\
${0.17}$ & ${-65.5}$ & ${3.0}$ & ${0.2}$ & ${18}$ & ${430}$ & ${600}$ 
& ${20}$\\
\noalign{\smallskip}
{\bf A\,2390}\\
${0.23}$ & ${0.0}$ & ${0.0}$ & ${0.1}$ & ${17}$ & ${1040}$ & ${900}$ 
& ${50}$\\
${0.23}$ & ${-0.04}$ & ${-2.70}$ & ${0.3}$ & ${18}$ & ${470}$ & ${60}$ 
& ${15}$\\
\noalign{\smallskip}
{\bf Cl\,2244$-$02}\\
${0.33}$ & ${0.0}$ & ${0.0}$ & ${0.17}$ & ${45}$ & ${680}$ & ${900}$ 
& ${30}$\\
${0.33}$ & ${17.32}$ & ${-10.2}$ & ${0.15}$ & ${80}$ & ${340}$ & ${600}$ 
& ${20}$\\
\noalign{\smallskip}
{\bf Cl\,0024+16}\\
${0.39}$ & ${0.3}$ & ${1.6}$ & ${0.3}$ & ${-10}$ & ${1030}$ & ${900}$ 
& ${25}$\\
${0.39}$ & ${1.78}$ & ${73.3}$ & ${0.3}$ & ${40}$ & ${240}$ & ${200}$ 
& ${15}$\\
\noalign{\smallskip}
{\bf Cl\,0054$-$27}\\
${0.57}$ & ${0.0}$ & ${-1.0}$ & ${0.2}$ & ${-22}$ & ${1100}$ & ${900}$ 
& ${30}$\\
\hline
\end{tabular}
\end{center}
\caption{Properties for the primary and 
(where relevant) secondary mass clumps in the
clusters. The characteristic parameters are initially constrained by 
the positions, shapes and luminosities of the multiple-imaged 
objects and are then iteratively varied to match the weak shear field
as well to obtain the optimal values (in the 
$\chi^2$ sense) in a likelihood scheme.}
\end{table*}

\subsection{Results from the lens models}

We successfully construct high resolution mass models for all five
clusters from the unambiguous galaxy-galaxy lensing signal detected
using the maximum-likelihood analysis. We constrain the mass enclosed
within an aperture for a fiducial halo. The maximum-likelihood
analysis yields the following: (i) the mass-to-light ratio in the
$V$-band of a typical $L^*$ galaxy does not evolve significantly as a
function of redshift, (ii) the fiducial truncation radius of an $L^*$
galaxy varies from about 20\,kpc to 70\,kpc depending on the cluster
(iii) the typical central velocity dispersion is roughly 180\,{\rm
\,km\,s}$^{-1}$. In Fig.~1, we show the recovered parameters
($\sigma_0*$, $r_t*$) for an $L^*$ galaxy from the maximum-likelihood
analysis. For the galaxy mass model adopted in our analysis, the total
mass of an $L^*$ galaxy varies from $\sim 4.9 \times 10^{11}\,{\rm
M}_{\odot}$ to $\sim 1.4 \times 10^{12}\,{\rm M}_{\odot}$. We note
that the distribution of these recovered parameters points to roughly
two kinds of typical haloes, compact, dense ones and more extended
ones. However, given that these lensing clusters do not constitute a
well defined sample it is premature to claim any dichotomy.  The
mass-to-light ratios obtained take passive evolution of elliptical
galaxies into account as given by the stellar population synthesis
models of Bruzual \& Charlot (2003), therefore any detected trend
reflects pure mass evolution. The mass obtained for a typical bright
cluster galaxy by Tyson et al.\ (1998) using only strong lensing
constraints inside the Einstein radius of the cluster Cl\,0024+1654,
at $z = 0.41$, is consistent with our results. All error bars quoted
here are $\sim 3\sigma$.

\begin{figure*}
\resizebox{10cm}{!}{\includegraphics[]{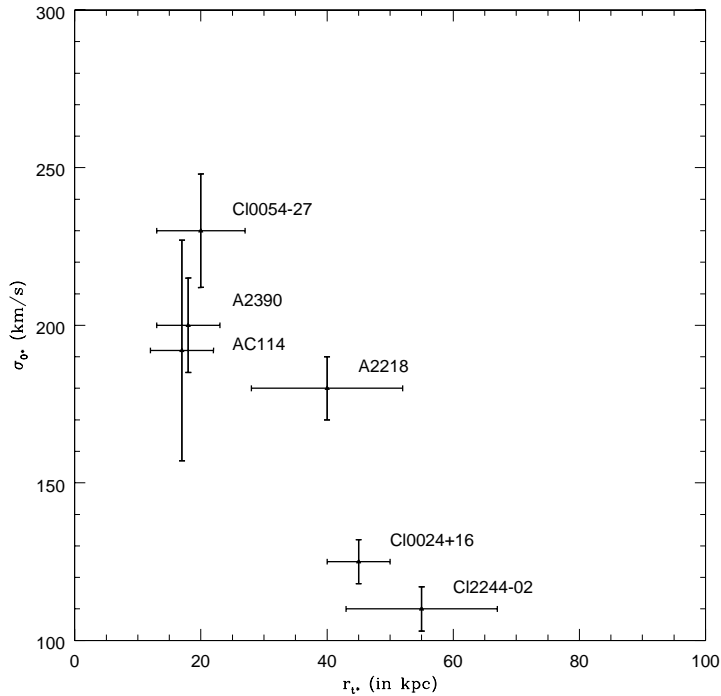}}
\caption{The retrieved best-fit values for the 2 parameters 
that were marginalised over in the likelihood analysis, the 
central velocity dispersion and aperture radius for an $L^*$ 
galaxy in each of the clusters. Included is the previously published 
data point for AC114. The error bars are 3$\sigma$ for both 
$\sigma_{0*}$ and $r_{t*}$. It is these error bars that translate
into a factor of 2 error in the total aperture mass.}
\end{figure*}

By construction, the maximum-likelihood technique presented here
provides the mass spectrum of subhaloes in the cluster directly. Note
that as stated before, in performing the likelihood analysis to obtain
characteristic parameters for the subhaloes in the cluster, it is
assumed that light traces mass. This is an assumption that is well
supported by galaxy-galaxy lensing studies in the field (Wilson et
al.\ 2001) as well as in clusters (Clowe \& Schneider 2002). In fact,
all lens modeling and rotation curve measurements suggest an excess of
baryons in the inner regions. Note however that for our choice of mass
model (the PIEMD) the mass to light ratio is not constant with radius
within an individual galaxy halo.  Since the procedure involves a
scaled, self-similar mass model that is parametric, we obtain a mass
estimate for the dark haloes (subhaloes) as a function of the
luminosity of the early-type galaxy hosted by them. This provides us
with a clump mass spectrum. We surmise that truncation by the cluster
causes these galaxy halo masses to be lower than that of equivalent
luminosity field galaxies at comparable redshifts obtained from
galaxy-galaxy lensing. The fraction of mass in these clumps is only
10--20 per cent of the total mass of the cluster within the inner
$500\,h^{-1}\,{\rm kpc}$ of these high central density clusters. The
remaining 80--90 per cent of the cluster mass is consistent with being
smoothly distributed (in clumps with mass $M\,<\,10^{10}\,{\rm
M}_{\odot}$, the precise composition of this component depends on the
hitherto unknown nature of dark matter. Note that the upper and lower
limits on the mass spectrum vary from cluster to cluster due to the
difference in the luminosity functions of cluster galaxies. These mass
functions and other detailed properties of the substructure can now be
directly compared to the subhalo mass functions of dark matter haloes
taken from the Millennium simulation, the results of which are
presented in Section~4.

\subsubsection{Uncertainties, sources of error and robustness of the lens
  models} 

The following tests were performed for each cluster, (i) choosing
random locations (instead of bright, early-type cluster member
locations) for the perturbers; (ii) scrambling the shapes of
background galaxies; (iii) choosing to associate the perturbers with
the 40 faintest (as opposed to the 40 brightest) galaxies; (iv)
randomly selecting known cluster galaxies as perturbers; (v) selecting
late-type galaxies. None of the above cases (i)--(v) yields a
convergent likelihood map, in fact all that is seen in the resultant
2-dimensional likelihood surfaces is noise. We averaged the shear
field in 4 radial bins around each cluster galaxy and stacked for each
cluster to estimate the aperture mass within those bins. We found that
the mean value of the shear in the inner-most radial bin was $\sim\,8\,
\pm\, 2$ \% falling off to $\sim $ 2\% in the last radial bin indicating
the presence of an aperture mass ranging from $10^{11} -
10^{12}\,\msun$. Choosing random locations in the cluster (i.e. not
the locations of early-type cluster members) we do not detect a shear
profile in the stacked, direct averaging.Alternatively, we also used the
locations of late-type galaxies (morphological classification from HST data 
is available for all the clusters studied here), to perform the direct averaging
and did not detect any shear.

The robustness of our results has been extensively tested, however
there are a couple of caveats that we ought to mention. As outlined
above, in this galaxy-galaxy lensing technique we are only sensitive
to a restricted mass range in terms of secure detection of
substructure.  This is due to the fact that we are quantifying a
differential signal above the average tangential shear induced by a
cluster, and we are inherently limited on by the average number of
distorted background galaxies that lie within the aperture scale radii
of cluster galaxies.  This trade-off between requiring a sufficient
number of lensed background galaxies in the vicinity of the subhaloes
and the optimum locations for the subhaloes leads us to choose the
brightest 40 early-type cluster galaxies for each lens. With deeper,
wider and more numerous images of clusters, expected in the future
with a wide-field imager in space such as the SNAP mission (Aldering
et al. 2003), this technique can be pushed much further to probe down
to lower masses in the mass function. It is possible that the bulk of
the mass in subhaloes is in lower mass clumps (which in this analysis
is essentially accounted for as part of the smooth component) and are
in fact anti-correlated with positions of early-type galaxies.

Our results still hold true since we are filtering out only the most
massive clumps via this technique. Note that one of the null tests
performed above, associating galaxy haloes with random positions in
the cluster (and not with the locations of bright, early-type
galaxies) resulted in pure noise with an average value of the measured
tangential shear ranging from $-0.10$ to $0.05$. Even if we suppose
that the bulk of the dark matter is associated with say, dwarf/very
low surface brightness galaxies in clusters, then the spatial
distribution of these galaxies is required to be fine-tuned such that
these effects do not show up in the shear field in the inner regions,
implying that if at all they are likely to be more significant
repositories of mass perhaps in the outskirts of clusters.

Guided by the current theoretical understanding of the assembly of
clusters, dwarf galaxies are unlikely to survive in the high density
core regions of galaxy clusters studied here. Studies such as the one
presented here when applied on larger scales to distances of a few Mpc
from the cluster centre will enable an understanding of the role of
environment in mass stripping of these haloes.  In a recent study,
Limousin et al. (2006) track the mass inferred for a fiducial cluster
galaxy as a function of cluster-centric distance out to twice the
virial radius. Using ground-based data they do not detect any
significant variation in mass, however, using mosaiced HST images 
Natarajan et al.(2006), find that a typical $L^*$ galaxy-halo has
higher aperture mass in the outskirts of the cluster. 

The principal sources of error in the above analysis are (i) shot
noise -- we are inherently limited by the finite number of sources
sampled within a few tidal radii of each cluster galaxy; (ii) the
spread in the intrinsic ellipticity distribution of the source
population; (iii) observational errors arising from uncertainties in
the measurement of ellipticities from the images for the faintest
objects and (iv) contamination by foreground galaxies mistaken as
background. As mentioned in Section 2.2.2, the partitioning of mass
into subhaloes and the smooth component as done here is largely
independent of the $N(z)$ of background galaxies.

In terms of the total contribution to the error budget, based on
simulations we find that the shot noise is the most significant source
of error, amounting to $\sim 50$ per cent; followed by the width of
the intrinsic ellipticity distribution which contributes $\sim 20$ per
cent, and the other three sources together contribute $\sim 30$ per
cent. This elucidates the future strategy for such analyses - going
significantly deeper and wider in terms of the field of view is likely
to provide considerable gains. Mosaic-ed ACS images are the ideal data
sets for this galaxy-galaxy lensing analyses, and such work is
currently in progress.
\section{Comparison with N-body simulations}

In this section, we present a comparison between the properties of the
subhaloes  determined using  the lensing  analysis detailed  above and
results from  N-body simulations. For this  study, we make  use of the
{\it  Millennium  Simulation},  recently  carried  out  by  the  Virgo
Consortium and  described in  detail in Springel  et al.  (2005).  The
simulation follows $N= 2160^3$ particles within a comoving box of size
$500\,   h^{-1}$Mpc   on   a   side.    With  a   particle   mass   of
$8.6\times10^{8}\,h^{-1}{\rm M}_{\odot}$  and a spatial  resolution of
$5\, h^{-1}$kpc,  the Millennium Simulation  provides an unprecedented
combination  of  high resolution  and  large  volume  that allows  the
formation  of rare objects  - like  massive lensing  clusters -  to be
followed  in a  representative  fashion.  It  therefore represents  an
ideal simulation  to compare with  observations as a  relatively large
sample of massive, lensing clusters can be extracted over the redshift
range probed by the observations. For the $64$ time slices produced by
the simulation, embedded substructures  within dark matter haloes were
identified  with  the algorithm  {\small  SUBFIND}  (Springel et  al.\
2001).   In brief,  the algorithm  decomposes a  given  particle group
(previously identified  with a standard  friends-of-friends algorithm)
into a set of disjoint substructures, each of which is identified as a
locally  overdense  region in  the  density  field  of the  background
halo. After  the regions containing substructure  candidates have been
identified, an  unbinding procedure is employed  to iteratively reject
all  particles  with positive  total  energy.  All substructures  that
survive  the  unbinding  procedure,  and  still  have  at  least  $20$
self-bound particles, are considered  to be genuine substructures.  We
refer to  the original  paper for more  details on the  algorithm.  We
note  however  that  the  identification of  subhaloes  within  haloes
represents  a difficult  technical  problem.  A  variety of  different
algorithms  have  been developed  in  order  to  accomplish this  task
(e.g. Gottl\"ober et al.\ 1998;  Klypin et al.\ 1999, Springel et al.\
2001; Weller et al. 2005), but unfortunately little work has been done
so  far  to  compare  the  properties  of  subhaloes  identified  with
different  methods, so  that  the systematic  effects and  differences
inherent in these methods are largely unknown.

In order to carry out our comparison with lensing results, we have
selected all haloes with $M_{200} \ge 8\times10^{14}\,{\rm M}_{\odot}$
from the simulation box at the snapshots corresponding to the
redshifts of the lensing clusters (see Table~1).  We find $17$, $15$,
and $12$ very massive haloes at the redshifts of the clusters A\,2218,
A\,2390, and Cl\,2244$-$02 respectively.  Massive cluster haloes are
on the tail of the mass function in a $\Lambda$CDM Universe and their
number decreases rapidly with increasing redshift. For Cl\,0024+16 and
Cl\,0054$-$27, we have therefore combined two adjacent snapshots (in
both cases the redshift of the observed clusters lies in between those
of the used snapshots), which led to $12$ and $6$ candidates,
respectively.
  
For each cluster halo, we have analysed the distribution of the
subhalo properties by projecting along the $x$, $y$, and $z$ axes and
keeping subhaloes $2\,{\rm Mpc}$ from the cluster centre that are
projected to within $1\,{\rm Mpc}$, identified with the position of
the most bound particle.  The distributions we discuss in the
following are then the average over the three projections of all the
haloes identified for each redshift.

\subsection{The mass function of substructures}

\begin{figure*}
\resizebox{18cm}{!}{\includegraphics[angle=0]{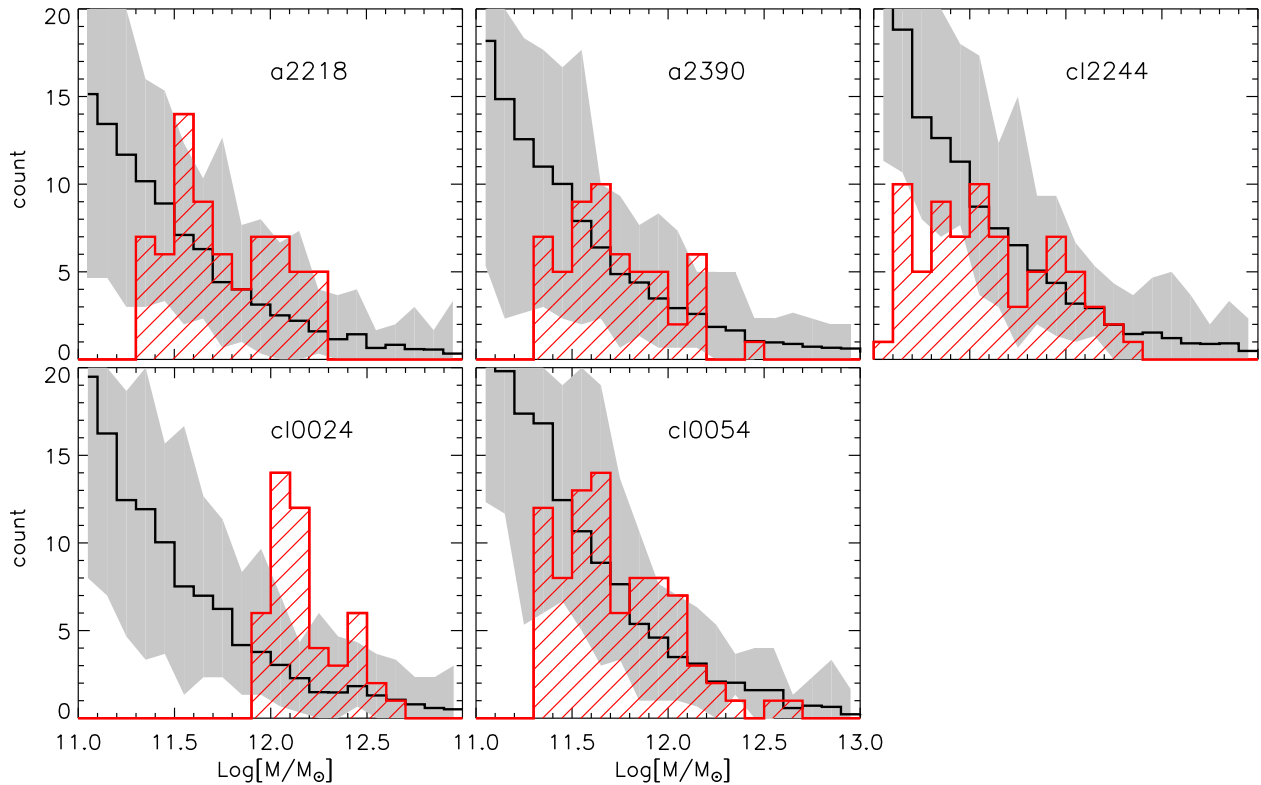}}
\caption{Comparison between substructure mass function retrieved from
  the galaxy-galaxy lensing analysis (red shaded histograms) and
  results from haloes selected from the Millennium Simulation.  The
  black solid line in each panel represents the average subhalo mass
  function of haloes selected at the redshift of the observed lensing
  cluster (see text for details).  The grey shaded region represents,
  for each value of the subhalo mass, the min-max number of
  substructures found in the simulated haloes. \label{Fig1}}
\end{figure*}

As mentioned in Sec.~1, a direct mapping of the substructure mass
function is ultimately of interest since it provides an important test
of the $\Lambda$CDM paradigm.  In addition, it is intimately connected
to the galaxy formation process: comparison of the dark halo mass
function with the observed luminosity function of galaxies can provide
valuable insights into the physical processes driving galaxy formation
and evolution. In this work, we restrict ourselves to a comparison of
the properties of the dark matter component. A yet more detailed
comparison for A\,2218 in terms of observable galaxy properties by
means of simulations and semi-analytic techniques is presented
elsewhere (De Lucia, Natarajan \& Springel 2007).

In Fig.~2 we compare the substructure mass function retrieved from the
galaxy-galaxy lensing analysis (hashed histograms) with the results
from the numerical simulation. The black solid line in each panel
represents the average over the three orthogonal projections of all
the massive haloes identified for each redshift. All subhaloes along
the line of sight that are projected to within 1 Mpc are included in
the inventory. The grey filled histograms show for each value of the
subhalo mass, the minimum and maximum number of substructures found in
the simulated clusters.  The lower mass cut-off for the mass function
determined from lensing is set by observational limitations, primarily
the accuracy with which anisotropies in the shear field can be
statistically detected.  We note that the mass of the subhaloes used
to build the grey histogram in Fig.~2 is defined in terms of the total
number of particles they contain, and {\em has been multiplied by a
correction factor} of $2$, as discussed below. As shown in Fig.~1, the
error bars in $\sigma_{0*}$ and $r_{t*}$ retrieved from the lensing
observations translate into a factor of 2 error in the aperture mass
of substructures.

When taken at face value this correction factor of 2 would signal a
discrepancy between the simulation and lensing results. However, we
argue that the difference is dominated by systematic effects in the
measurement of substructure masses in the simulation, and possibly
also in the lensing analysis.  Note that the algorithm {\small
SUBFIND} identifies substructures as overdense regions bounded by an
isodensity surface where the density equals the local value of the
background host halo.  Because of this criterion for identifying
substructure, it is possible that our subhalo masses are biased low,
because only the `tip of the iceberg' is seen. This effect should
increase the closer a subhalo is located to the centre. We will now
show that this suspicion is indeed true.

For our comparison with lensing data, the most important question is
whether the self-bound mass reported by {\small SUBFIND} for a
substructure actually accounts for the {\em full} mass enhancement
present at the location of the substructure in the N-body simulation.
To determine this mass enhancement, one can measure the enclosed mass
within a little sphere around the location of the substructure and
subtract from it the mass in the same sphere after spherically
averaging the whole mass distribution of the halo around the halo
centre. The latter gives an estimate of the background density in the
volume occupied by the substructure. If the mass distribution is a
superposition of a spherically symmetric background halo and an
embedded substructure, then this procedure will (almost exactly)
recover the true mass of the substructure independent of the density
profile of the background halo and that of the substructure, provided
that the spherical aperture fully encloses the substructure. The mass
estimate will be biased slightly low because the spherically averaged
mass distribution includes the substructure mass, but this should be
negligible if the volume of the substructure is small against that of
the background halo.

This leaves the question of what to choose for the radius of the
sphere that is needed in the measurement. If there was only a single
substructure in a spherically symmetric halo, the result would be
invariant once the sphere is large enough to fully enclose the
substructure. In real-world haloes we will suffer from growing errors
in the measurement when the aperture becomes too large, both from
other nearby substructures and from the fact that the background halo
is not exactly spherical in shape. As a simple compromise we have
adopted 3 times the half-mass radius measured for the substructure
(the half-mass radius is the radius that enclosed half the bound
particles) to carry out our measurements of the mass
enhancements. Note that for a NFW halo, the virial radius would be 2.8
times the half-mass radius for a concentration of 10, while it would
be 3.2 times the half-mass radius for a concentration of 15.

\begin{figure}
\begin{center}
\resizebox{8.5cm}{!}{\includegraphics{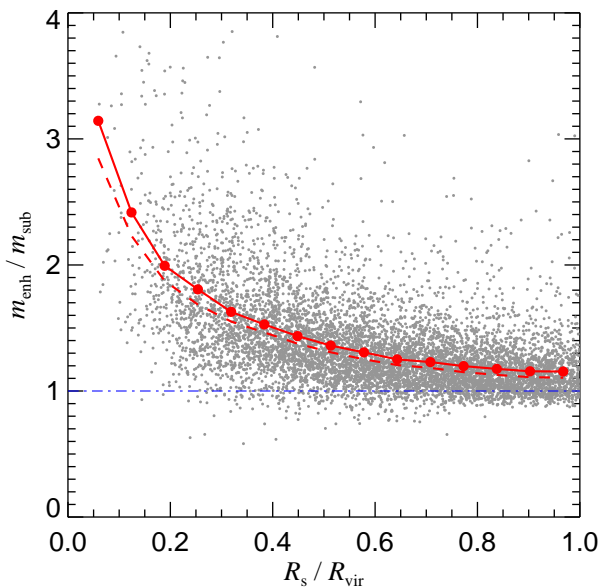}}
\end{center}
\caption{Comparison of the actual mass enhancements found around the
locations of substructures ($m_{\rm enh}$) with the self-bound mass
$m_{\rm sub}$ reported by {\small SUBFIND}, as a function of radial
distance $R_s$ in their parent halo of virial radius $R_{\rm vir}$.
All substructures with more than 1000 particles in haloes above
$10^{14}\,h^{-1}{\rm M}_\odot$ in the Millennium were analysed. A
random 25\%-subsample of the points are included in the scatter
diagram.  The solid-line with symbols gives the mean in each radial
bin for the whole sample, while the dashed line is the median in each
bin.
 \label{FigMassBias}}
\end{figure}

Adopting the above measurement procedure, we have determined the mass
enhancements $m_{\rm enh}$ around the positions of all substructures
identified by {\small SUBFIND} in haloes with masses above
$10^{14}\,h^{-1}{\rm M}_\odot$, and with a minimum particle number of
1000 (in total $\sim 31000$ objects).

In Fig.~\ref{FigMassBias}, we show the ratio between the measured
mass-enhancement and the corresponding {\small SUBFIND} self-bound
mass as a function of radial position of the substructure within the
parent halo.  There is clearly a mass-bias with a strong radial
dependence. The $m_{\rm sub}$ masses reported by {\small SUBFIND} for
substructures in the inner parts of clusters are systematically biased
low; the real mass enhancements relevant for lensing can be a factor
2-3 higher in the innermost parts of the cluster. For the bulk of the
substructures at larger radii, there is only a small difference of
order $\sim ~20\%$ or so, a value that could also be weakly affected
by our choice of aperture radius. However, the radial trend of the
mass ratios is robust. While most of the overall substructure mass is
found in the outer parts of the cluster, we note that the luminous
cluster galaxies are much more centrally concentrated (Springel et
al. 2001), so that the bias in the inner parts enters with a large
weight in our comparison of lensing data with the N-body simulations.
This provides the quantitative basis for our fiducial correction factor
of 2 that we applied in the comparison shown in Fig.~2. We think a
factor of this order is plausible given the systematics of current
techniques, but it is also clear that a more precise determination
will require an improved understanding of the systematics of both the
substructure finder and the lensing analysis.

We note that we have also checked whether there is any mass dependence
of this `bias' in the {\small SUBFIND} masses, but we have found no
evidence for this.  This is consistent with the fact that there is a
good uniformity of the measured slopes for the subhalo mass function
from different studies (De Lucia et al.\ 2004; Diemand et al.\ 2004),
which suggests that the systematics of the substructure finders can be
accounted for through simple scale factors, at least at a statistical
level.  Also, we think that it is likely that other algorithms for
identifying substructures suffer from this problem in similar ways.

In addition to the above, is is likely that there are some systematic
effects in the mass estimate based on the lensing analysis as
well. Here a systematic overestimate appears more likely, arising for
example if the most massive substructures correspond to haloes that
have fallen in most recently, as indicated by recent numerical studies
(De Lucia et al.\ 2004; Gao et al.\ 2004). As for the mass
reconstruction technique employed here, biases can also be introduced
due to the possible existence of substructure along the line of sight,
not in the lens plane but behind it (Natarajan 2006). Fig.~2 also       
shows that there is a large scatter among simulated clusters and since
the number of substructures in the observable mass range is quite
small, we also expect large system-to-system variations between
different observed systems. Besides, as shown in Table~1, the subhalo
masses are uncertain to within a factor of 2 from the galaxy-galaxy
lensing analysis, based on the observational limitations from current
data.

We note here in our preliminary comparison of the lensing derived mass
function with that extracted from the simulation of a single, massive
cluster (Natarajan \& Springel 2004), we had found excellent agreement
without the need for a mass correction factor. However, it turns out
that this result was influenced by not correctly taking into account
the spatial geometry of the HST field. When this effect is
appropriately included, our present results are consistent with this
earlier work, and a mass correction factor of 2 is required to obtain
quantitative consistency. Note that this is within the limits of the
systematic uncertainties both of the substructure mass detection in
the simulations as well as those from the lensing analysis. We appear to 
be missing subhaloes at the high mass end ($log M \geq 12.5$), in the 
lensing determination. This is likely due to the fact that such high 
mass haloes probably host more than one luminous component. We plan to
investigate this issue further in our detailed study of A2218 in future
work.

Overall, it is therefore remarkable that we find very good qualitative
agreement in the mass range sampled by the lensing analysis, and very
good quantitative agreement as well when the systematic biases are
approximately corrected for within the errors.  It is interesting to
note here that the only obvious outlier is Cl\,0024+16.  This is the only
system that clearly has a bimodal structure with two almost equal mass
clumps separated by a small redshift offset. Its bimodal structure has
been interpreted as a sign of a recent merger (Czoske et al. 2002).
Clearly, if this is indeed the case, the comparison with relaxed
massive haloes from the Millennium Simulation is not appropriate.  We
plan to investigate this question in more detail in future work by
comparing the properties of the substructures in Cl\,0024+16 with clusters
selected from the Millennium Simulation that are recent mergers of
equivalent mass.

\subsection{The distribution of velocity dispersions}

\begin{figure*}
\resizebox{18cm}{!}{\includegraphics[angle=90]{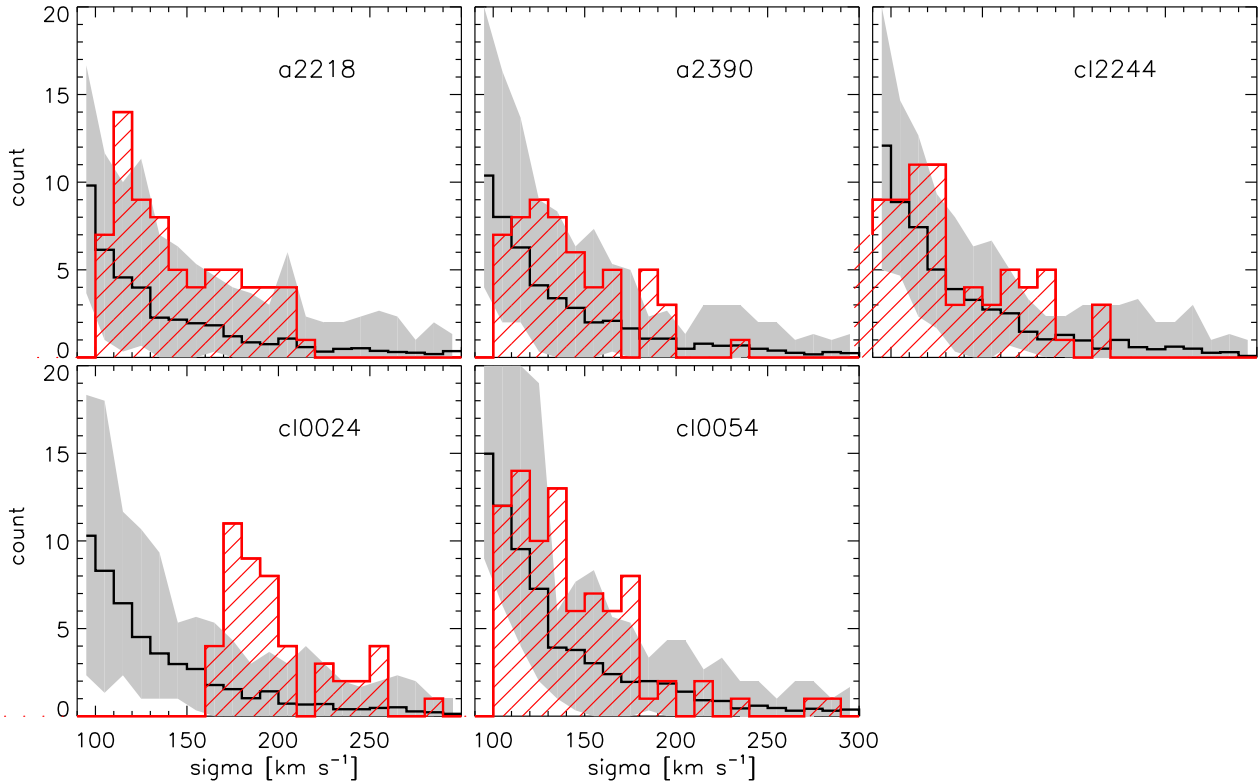}}
\caption{As in Fig.~2 but for the distribution of velocity
dispersions.}
\end{figure*}

In Fig.~4, we compare the distribution of velocity dispersions
retrieved by the lensing analysis and the average distributions
obtained for the simulated clusters selected from the Millennium
Simulation.  The velocity dispersions of substructures in simulated
clusters are estimated using the velocity information for all the
particles attached to the dark matter substructure and have been
scaled by a factor of $\sqrt{2}$ (see previous section).  Again we
find generally a good agreement over the range sampled by the lensing
analysis, except for Cl\,0024+16, which yet again appears to be an outlier
(see discussion in previous section).  Note here that the velocity
dispersion retrieved from the lensing analysis is one of the two
parameters that characterises each subhalo. This velocity dispersion
is the normalization of the Faber-Jackson relation, therefore, relates
the mass (for the PIEMD model) to that of early-type galaxies. For the
simulated haloes the measured velocity dispersion is that of the dark
matter alone.  We note that Springel et al. (2001) used a factor equal
to $0.9$ to convert the 1D dispersions of subhaloes into stellar
velocity dispersions, which was adopted to match the zero-point of the
Faber-Jackson relation.  Due to the dissipation involved in star
formation, it is plausible that the stellar velocity dispersion is
{\it smaller} than that of the dark matter.  On the other hand, we
argued above that our mass estimates might be biased low. We explore
spatial and velocity biases in more detail using the lensing results
from A\,2218 in a forthcoming paper.

\subsection{The distribution function of tidal radii}

\begin{figure*}
\resizebox{18cm}{!}{\includegraphics[angle=90]{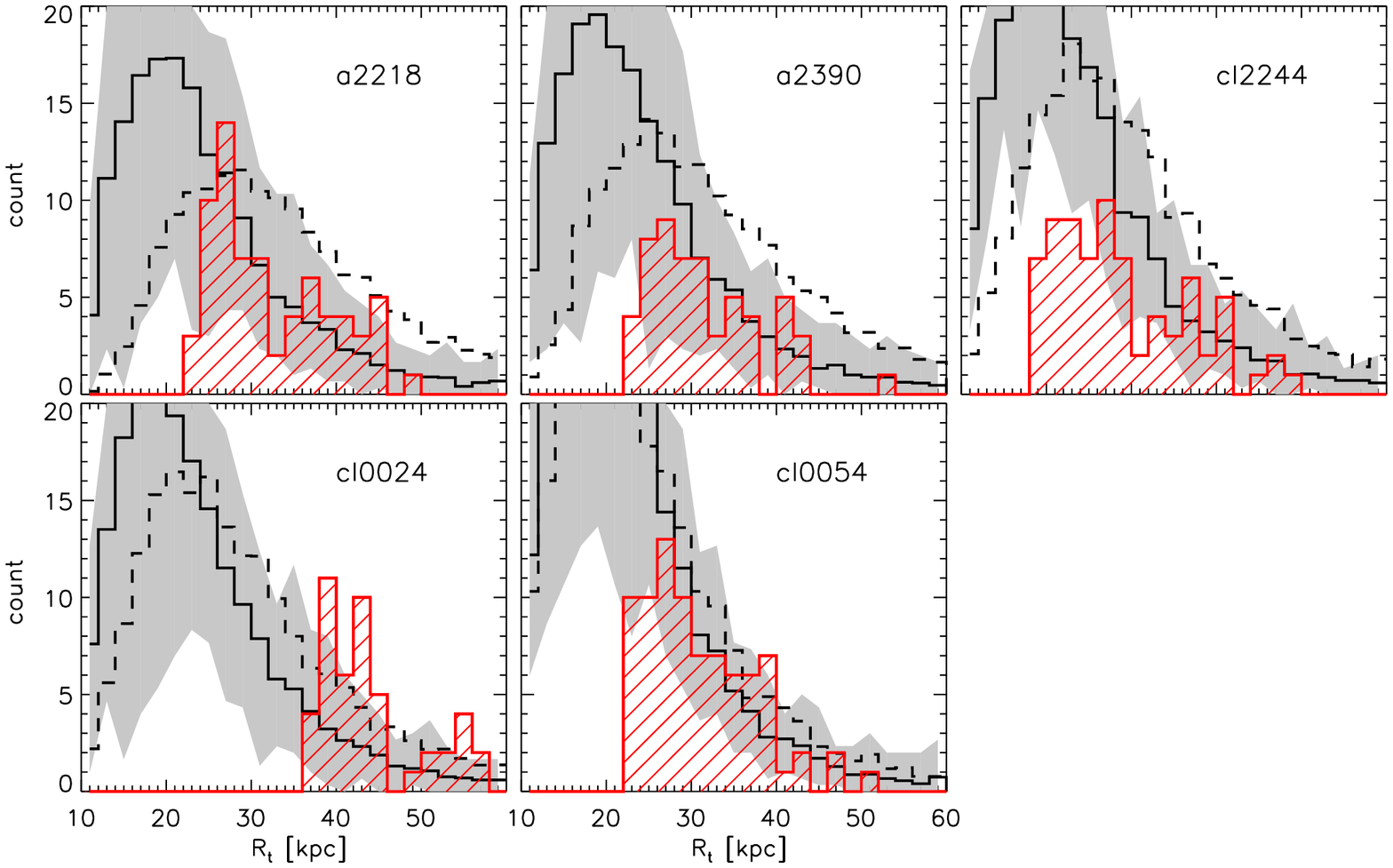}}
\caption{As in Fig.~2 but for the distribution of tidal
  radii.  The solid black line with the grey regions are obtained by
  measuring for each substructure, the radius corresponding to its
  half mass.  The dashed black line is obtained by taking the radius
  corresponding to the maximum circular velocity.}
\end{figure*}

We use the aperture radii derived from the lensing analysis as a proxy
for the tidal radii for the subhaloes. These aperture radii are more
compact than the tidally truncated extents of the subhaloes. This
radius cannot be identified directly with the tidal radius defined in
the usual sense, but is proportional to it.  In Fig.~5, we finally
show the comparison between the distribution of aperture radii ($r_t$)
retrieved by the lensing analysis with the average distributions of
the tidal radii of substructures in simulated clusters selected from
the Millennium Simulation.  The solid black line in the figure and the
grey filled region are obtained by measuring, for each substructure,
the radius $r_{\rm half}$ that divides the halo mass in equal inner
and outer parts.  We note that, by using this measure of the radius,
we find a very nice agreement with the distributions retrieved by the
lensing analysis (with the usual exception of Cl\,0024+16) over the full
range sampled by the observations.  $r_{\rm half}$, however,
represents an underestimate of the `tidal radius' of a dark matter
substructure.  The dashed black line in Fig.~5 gives the distribution
obtained by using the radius corresponding to the maximum circular
velocity ($r_{\rm max}$), which is offset with respect to $r_{\rm
half}$ by a factor that is slightly less than $2$.  We note that, as
explained in the previous sections, the mass retrieved by the lensing
analysis is essentially the mass within a fixed aperture, that is then
identified as being proportional to the tidal radius of the associated
dark matter substructure.  Given the uncertainty in the measurement
technique, the good agreement shown in Fig.~5 is therefore noteworthy.
We recall that, in order to obtain a good agreement with the observed
subhalo mass function, we have corrected their mass by a factor of
$2$.  We are therefore in a situation where the substructures detected
in the simulations are slightly bigger but less massive than those
retrieved by our lensing technique.  Such a situation would be
expected if the density profiles assumed for the substructures are
systematically denser in the inner parts than the simulated subhaloes.
And, in fact, recent numerical studies suggest that dark matter
subhaloes have softer profiles than NFW (Hayashi et al.\ 2004; Stoehr
et al.\ 2003).  Unfortunately our lensing technique is not sensitive
to the choice of mass profile, rather it represents a tool to
determine the total mass enclosed within an aperture.  Higher
resolution data will in the future allow the slopes of the mass
profiles in substructure to be constrained. Adiabatic contraction 
of the dark matter density profile in response to the baryonic
component in the inner regions of the cluster halo (on scales of 
$r < 10 - 50$ kpc) is also expected to be relevant, however the importance
and significance remains to be assessed from numerical simulations. There
are two conflicting claims on how significant this effect at the present
time (Zappacosta et al. 2006; Rozo, Nagai, Keeton \& Kravtsov 2006).

\section{Conclusions and Discussion}

In this paper, we present (i) high resolution mass models for lensing
clusters, (ii) the inferred mass function of subhaloes inside these
clusters and (iii) a detailed comparison of the subhalo mass function,
the velocity dispersion and aperture radii function with an ensemble
of cluster--sized haloes selected from the Millennium
Simulation. Detailed results of the application of our galaxy-galaxy
lensing analysis techniques to five HST cluster lenses are used to
construct high resolution mass models of the inner regions. In order
to do so we have utilised both strong and weak lensing observations
for these massive clusters.  The goal has been to quantify
substructure in the cluster assuming that the subhaloes follow the
distribution of bright, early-type cluster galaxies.  Similar attempts
have been made in the lower density field environment yielding typical
galaxy masses and central velocity dispersions. The mass distribution
for a typical galaxy halo inferred from field studies is extended with
no discernible cut-off in most analyses with the exception of one
study by Hoekstra et al.(2004) where they report $r_t =
260^{+124}_{-73}$ kpc using data from the CNOC-2 fields. By contrast,
in the cluster environments probed in this work we detect an edge to
the mass distribution in cluster galaxies. We have performed various
stringent checks to ascertain that this is not an artifact of the
choice of mass model.  Rather this result provides evidence for tidal
stripping by the global cluster potential well.

Aside from the detailed lens models, we also present the mass spectrum
(albeit within a limited mass range with subhalo masses ranging from
$10^{11} - 10^{12.5}\, \msun$) of substructure in the inner regions of
these clusters.  The survival and evolution of substructure offers a
stringent test of structure formation models within the CDM paradigm.
Subhaloes of the scale detected in all these clusters indicate a
significant probability of galaxy--galaxy collisions over a Hubble
time within a rich cluster.  However, since the internal velocity
dispersions of these clumps associated with early-type cluster
galaxies ($\sim 150$--250\,{\rm km\,s}$^{-1}$) are much smaller than
their orbital velocities, these interactions are unlikely to lead to
mergers, suggesting that the encounters of the kind simulated in the
galaxy harassment picture by Moore et al.\ (1996) are the most
frequent and likely.  High resolution cosmological N-body simulations
of cluster formation and evolution (De Lucia et al.\ 2004; Ghigna et
al.\ 1998; Moore et al.\ 1996), find that the dominant interactions
are between the global cluster tidal field and individual galaxies
after $z = 2$. The cluster tidal field tidally strips galaxy haloes in
the inner 0.5 Mpc efficiently and the radial extent of the surviving
haloes is a strong function of their distance from the cluster
centre. Much of this modification is found to occur between $z =0$ and
$z=0.5$.

We have performed a detailed comparison of the subhalo mass function,
velocity dispersion function and the distribution of aperture radii
retrieved from lensing with an ensemble of massive clusters from the
Millennium Simulation, which provides an ideal data-set for such an
analysis. A series of systematic effects, due to the algorithm used to
identify substructures in the simulation, needs to be taken into
account when performing such a comparison.  In addition, it should be
kept in mind that the level of uncertainty of the observational
results is still relatively high (for instance the mass derived from
lensing is only accurate to within a factor of two and the technique
is not sensitive to the choice of the substructure mass profile).
Overall, we find consistency between the distribution of substructure
properties retrieved using the lensing analysis and those obtained
from simulations, although our detailed comparison seems to suggest
that simulated substructures are slightly bigger but less massive than
sub-clumps detected by means of lensing techniques.  This might be due
to systematic differences between the density profiles of simulated
substructures and those assumed in the lensing model.  Unfortunately,
the technique is not sensitive to this choice but higher resolution
data will allow in the future the slopes of the mass profiles in
substructures to be constrained.

Despite the uncertainties mentioned above, the general agreement
between simulations and results determined {\it independently} from
lensing is remarkable. Our work represents a powerful test of the
$\Lambda$CDM model, which at present appears to be consistent with the
amount of observed substructure in massive, lensing clusters, up to
redshifts of $\sim 0.6$, given the uncertainties. It will be very
interesting to tighten the constraints with future lensing data.

\section*{Acknowledgements}

The authors acknowledge Simon White for support and encouragement
throughout this project. PN is grateful to the Virgo Consortium for
access to the Millennium Simulation data.  She acknowledges support
from NASA via HST grant HST-GO-09722.06-A. She also thanks her
collaborators Jean-Paul Kneib, Ian Smail and Richard Ellis for help
with the observational data and useful input on the work. GDL thanks
the Alexander von Humboldt Foundation, the Federal Ministry of
Education and Research, and the Programme for Investment in the Future
(ZIP) of the German Government for financial support.

\end{document}